%% file: npdgamma_lattice_021813.tex
\documentclass[prl,showpacs,showkeys,aps,amsfonts,amssymb,byrevtex,pdftex ,twocolumn ]{revtex4-1}
\pdfoutput=1

\usepackage{graphicx}  
\usepackage{amsmath, amssymb, bm}   
\usepackage[dvips]{color}
\usepackage{natbib}
\usepackage{hyperref}

\def\){\right)} 
\def\({\left(} 
\def\]{\right]} 
\def\[{\left[}

\begin{document}
\preprint{INT-PUB-13-007}

\title{
Radiative capture reactions in lattice effective field theory}

\author{%
Gautam Rupak}
\email{grupak@u.washington.edu}

\affiliation{Department of Physics $\&$ Astronomy, 
Mississippi State
University, Mississippi State, MS 39762, U.S.A. }

\author{%
Dean Lee}
\email{dean\_lee@ncsu.edu}

\affiliation{Department of Physics,
North Carolina State University, Raleigh, NC 27695, U.S.A.}

\begin{abstract}

We outline a general method for computing nuclear capture reactions on the lattice.  The method consists of two major parts.  In this study we detail the second part which consists of calculating an effective two-body capture reaction on the lattice at finite volume.  We solve this problem by calculating the two-point Green's function using an infrared regulator and the capture amplitude to a two-body bound state.  We demonstrate the details of this method by calculating on the lattice the leading M1 contribution to the radiative neutron capture on proton at low energies using pionless effective field theory.  We find good agreement with exact continuum results.

\end{abstract}



\maketitle

\emph{Introduction.}--- 
A long standing goal of nuclear physics and astrophysics is to understand
the nuclear reactions that made the elements of nature, 
from the lightest nuclei in the early universe to the light, medium, 
and heavy nuclei synthesized in stars.  In the past few years there has been progress in 
\textit{ab initio} calculations of nuclear reactions which goes beyond 
transitions from one bound state to another.  Recent work includes calculations 
using the No-Core Shell Model and resonating group method
\cite{Navratil:2011sa,Navratil:2011zs}, Fermionic Molecular
Dynamics \cite{Neff:2010nm}, the coupled-cluster approach
\cite{Jensen:2010vj}, and variational Monte Carlo \cite{Nollett:2011qf}.  
There has also been some progress using lattice calculations in finite periodic 
volumes to analyze coupled-channel scattering
\cite{Lage:2009zv,Bernard:2010fp,Meyer:2012wk,Briceno:2012yi} and three-body systems
\cite{Polejaeva:2012ut}.
However there is no general formalism for calculating inclusive and exclusive 
reactions from lattice simulations.  In this letter we present the first steps 
towards a general method for computing nuclear reactions on 
the lattice.  Our discussion here focuses on radiative capture reactions using 
the formalism of lattice effective field theory.  Our results will also have direct 
applications to ultracold atomic systems where there are interesting  
phenomena analogous to nuclear capture reactions.  In this case radio-frequency 
photons are used to induce the association of weakly-bound molecules 
\cite{Chin:2005a,Hanna:2007a,Weber:2008a,Machtey:2012a,Bazak:2012a}.

Lattice effective field theory combines the theoretical framework of effective
field theory (EFT) with numerical lattice methods.  A review of lattice effective 
field theory calculations can be found in Ref.~\cite{Lee:2008fa}. The method has been
applied to nuclei in pionless EFT \cite{Borasoy:2005yc} and chiral EFT
\cite{Epelbaum:2009pd,Epelbaum:2010xt,Epelbaum:2011md}. In Ref.~\cite{Epelbaum:2012qn} a new technique was developed which allowed for
general wavefunctions to be used as initial and final states in the lattice
EFT calculations.  It has since been realized that cluster wavefunctions can
be used to study the scattering and reactions of continuum states.  The
general strategy involves separating the calculation into two parts. The
first part of the method is to use projection Monte Carlo to determine a
multi-channel adiabatic lattice Hamiltonian for the participating nuclei.  For example, let
us consider a two-body capture reaction involving three nuclei,
$1+2\rightarrow3+\gamma$.  Let $\left\vert \vec{r}_{12}\right\rangle $ be a
initial cluster wavefunction with separation vector $\vec{r}_{12}$ between
nuclei $1$ and $2$. \ We consider all possible values for $\vec{r}_{12}$ and
use Euclidean time propagation to construct projected states, %
$
\left\vert \vec{r}_{12}\right\rangle _{t}=\exp\left(  -Ht\right)  \left\vert
\vec{r}_{12}\right\rangle$. 
For large $t$, the set of states $\left\vert \vec{r}_{12}\right\rangle _{t}$
will approximately span the linear space of continuum states for nuclei $1$
and $2$.  Using these states we can calculate one-photon transition matrix
elements with nucleus $3$.  The details of this projection technique and
construction of the multi-channel adiabatic lattice Hamiltonian will be discussed in
forthcoming publications.
The focus of this letter is describing how to do the second part of the reaction 
calculation. This second part uses the adiabatic lattice Hamiltonian for the participating
nuclei to calculate nuclear reaction rates.  For the case of nucleon-nucleus and 
nucleus-nucleus capture reactions, the problem is equivalent to the radiative capture of two
point particles forming a two-body bound state.  In this letter
we show how this calculation is done by using the well-known example of
radiative capture of a proton and neutron to form a deuteron. In our
analysis we use pionless EFT at leading order. We will show
that the continuum results which have been obtained in
Ref.~\cite{Chen:1999bg,Rupak:1999rk} can be accurately 
reproduced on the lattice.

\emph{Neutron capture.}---
The radiative capture process $p(n,\gamma)d$ is important in nuclear physics as it provides stringent bounds on the primordial deuterium abundance.  This abundance is sensitive to the amount of baryonic matter in the universe.  At very low energies, this reaction proceeds primarily through the M1 transition from the incoming $p+n$ spin-singlet $s$-wave state to the final spin-triplet $s$-wave deuteron bound state.  It is known that these initial and final $s$-wave states are non-perturbative.  Therefore calculating the M1 transition using lattice EFT provides a non-trivial check of the method developed here.  We add that at the energies relevant in Big Bang nucleosynthesis, $p(n,\gamma)d$ is dominated by both the M1 and E1 transitions. However, the initial state $p$-wave interaction associated with the E1 transition is perturbative.  So while we do not include the E1 contribution in the lattice EFT calculation described in the following, this contribution can be added trivially. 

The dominant M1 contribution can be calculated with the leading-order Lagrangian,
\begin{align}\label{eq:Lagrangian}
\mathcal L = & N^\dagger[iD_0+\frac{D^2}{2M}]N +\frac{e\kappa_1}{2 M}N^\dagger\tau_3\bm{\sigma}\cdot\bm{B} N \nonumber\\
-&\frac{c_s}{8}\sum_{i=1}^3( N\sigma_2\sigma_i\tau_2 N)^\dagger( N\sigma_2\sigma_i\tau_2 N)\nonumber\\
-&\frac{c_t}{8}\sum_{i=1}^3
( N\sigma_2\tau_2\tau_i N)^\dagger( N\sigma_2\tau_2\tau_i N)\, ,
\end{align}
where $D^\mu=\partial^\mu+ie(1+\tau_3) A^\mu/2$ is the covariant derivative, $B_i$ the magnetic field and $\kappa_1=2.35$ the isovector nucleon magnetic moment.  We take the nucleon mass as $M=939$ MeV. 
The Pauli matrices $\sigma_i$ act on the spin indices and $\tau_i$ act on the isospin indices of the nucleon field $N$. The Pauli matrices $\sigma_2\tau_2\tau_i$  and   $\sigma_2\sigma_i\tau_2$   project the incoming and final state nucleons onto the spin-singlet and spin-triplet channels, respectively.

\begin{figure}[thb]
\begin{center}
\includegraphics[width=0.45\textwidth,clip=true]{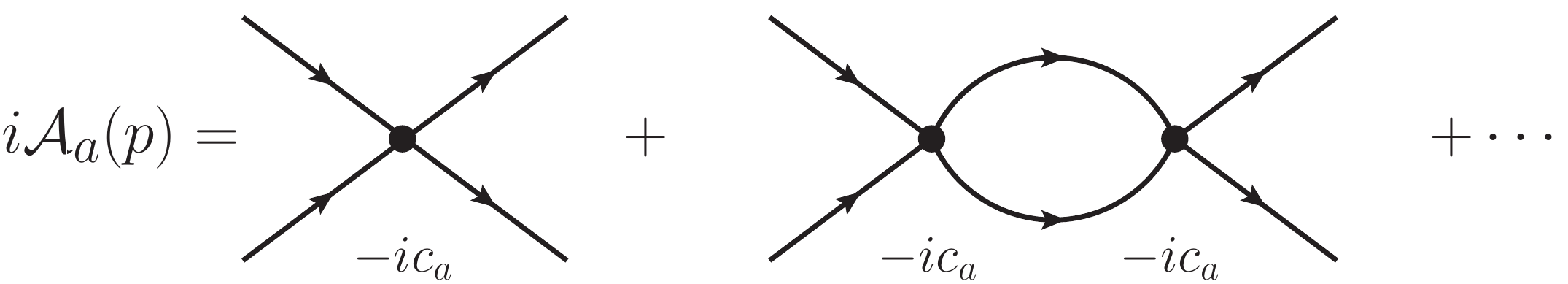} 
\end{center}
\caption{\protect Elastic scattering of the nucleon.  The index $a=s,t$ corresponds to the singlet and triplet channels.  
The ``$\cdots$" represents iteration of the interaction.}
\label{fig:scattering}
\end{figure}
The elastic scattering diagrams in Fig.~\ref{fig:scattering} give the amplitude
\begin{align}
iA_a(p)
=&i\frac{4\pi}{M}\frac{1}{-4\pi /(M c_a)-\lambda-ip},
\end{align}
where $\lambda$ is the renormalization scale and the index $a=s,t$ corresponds to the spin singlet and triplet channels, respectively.  In the spin-singlet channel, we tune the EFT couplings in the continuum such that there is a shallow virtual state with a scattering length $a= -23.7$ fm with $4\pi/(M c_s)+\lambda = 1/a$. In the spin-triplet channel we tune the EFT couplings to form a shallow bound state with binding momentum $\gamma = 45.7$ MeV  where $4\pi/(M c_t)+\lambda = \gamma$.

\begin{figure}[thb]
\begin{center}
\includegraphics[width=0.35\textwidth,clip=true]{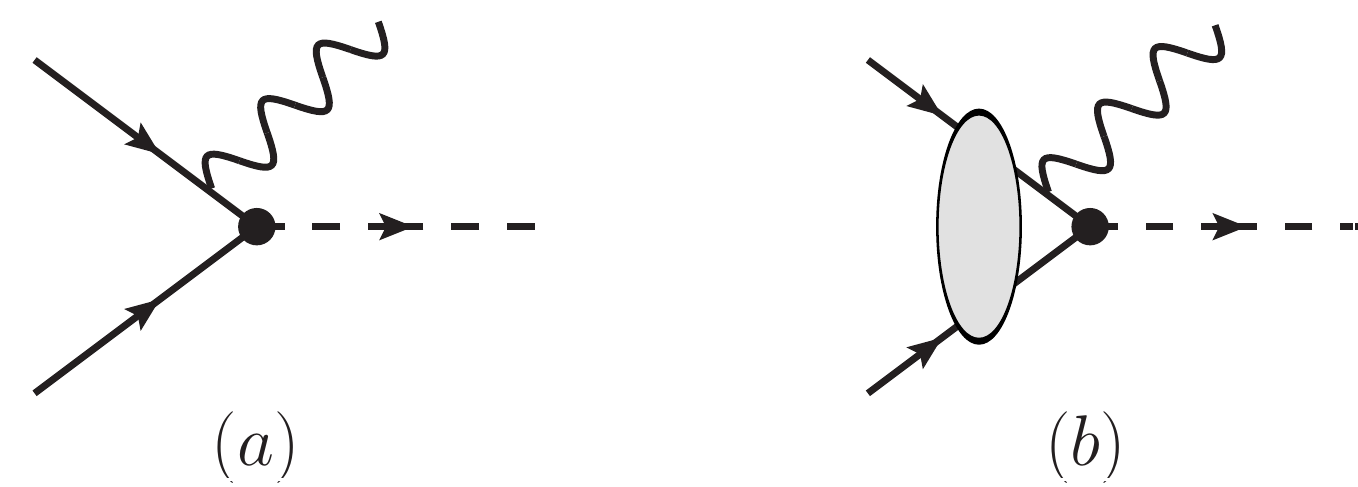} 
\end{center}
\caption{\protect Radiative capture diagrams for $p(n,\gamma)d$ .  Wavy line represents the photon, dashed line the deuteron, and blob the set of possible initial state interactions.  }
\label{fig:E1Capture}
\end{figure}
 The leading order contribution to the capture amplitude comes from the diagrams $(a)$ and $(b)$ in Fig.~\ref{fig:E1Capture}.   Using the center-of-mass (c.m.) kinematics with $\bm{p}$ the neutron momentum and $\bm{k}$ the photon momentum, we get~\cite{Chen:1999bg,Rupak:1999rk} in the continuum 
\begin{align}\label{eq:EFT}
(a)+(b)=&\frac{e\kappa_1}{2}\sqrt{2}\sqrt{2M}\sqrt{Z}\bm{\epsilon}_d^\ast\cdot(\bm{k}\times{\bm\epsilon}^\ast_\gamma)\nonumber\\
&\times U_N^T\sigma_2\tau_2\tau_3U_N \mathcal M_C,\nonumber\\
\mathcal M_C=&  \frac{1}{p^2+\gamma^2}-\frac{1}{(1/a+ip)(\gamma-ip)}.
\end{align}
$U_N$ represents the spinor wave functions for the nucleon fields, and $Z$ is the wave function renormalization factor defined as the residue at the deuteron propagator pole.  In the lattice calculation, we will reproduce the reduced matrix amplitude $\mathcal M_C$, and not concern ourselves with the overall factors associated with the standard normalization of non-relativistic cross sections.

The amplitude $\mathcal M_C$ is related to the expectation value $\langle\psi_B| O_\mathrm{EM}|\psi_i\rangle$ of the relevant  electromagnetic operator between final bound state wave function $\psi_B(\bm{r})$ and initial state  wave function  $\psi_i(\bm{r})$. 
On the lattice, it is straightforward to calculate the bound state wave function $\psi_B(\bm{r})$  from the discretized Hamiltonian.  The incoming wave function $\psi_i(\bm{r})$ is problematic since we work with periodic finite volumes.  One could do a finite volume analysis as in Ref.~\cite{Meyer:2012wk}, but this mixes in the elastic phase shift with the capture amplitude. Here instead we calculate the inelastic process on the lattice using the interacting two-point retarded Green's function. We write, in lattice units,
\begin{align}\label{eq:Greens}
&\mathcal M(\epsilon)=(\frac{p^2}{M} -E -i\epsilon) 
\sum_{\bm{x}, \bm{y}} \psi_B^\ast(\bm{y})  
\, G (E;\bm{x},\bm{y}) e^{i\bm{p}\cdot \bm{x}}
, \nonumber \\
&G(E;\bm{x},\bm{y})= \langle \bm{y}|\frac{1}{E -\hat{H}_s+i\epsilon} |\bm{x}\rangle,
\end{align}
where $G(E;\bm{x},\bm{y})$ is the Green's function for propagation from $\bm{x}$ to $\bm{y}$ with the strong interaction Hamiltonian $\hat{H}_s$ in the incoming spin-singlet channel. 
The bound state wave function is normalized such that 
$\sum_{\bm{x}}|\psi_B(\bm{x})|^2=1/(8\pi\gamma)$.  For on-shell incoming particles, the pre-factor 
$\frac{p^2}{M} -E -i\epsilon$ is just  $-i\epsilon$ in 
Eq.~(\ref{eq:Greens}), as prescribed by the Lehmann-Symanzik-Zimmermann reduction formula for the scattering matrix.   

For the lattice calculation of $\mathcal M$ in Eq.~(\ref{eq:Greens}) we use the Hamiltonian lattice formalism with continuous time.  The lattice Hamiltonian is derived from discretizing the continuum Hamiltonian corresponding with Eq.~(\ref{eq:Lagrangian}).  We use the simplest possible discretization with nearest neighbor hopping terms for the kinetic energy term and single-site two-body contact interactions.   We use a lattice spacing of $b = 1/100~$MeV$^{-1}$ for scattering momenta $p < 40~$MeV and $b = 1/200~$MeV$^{-1}$ for scattering momenta $p > 40~$MeV.  In each case we use periodic boundaries in all spatial directions and consider cubic lengths up to $L = 240b$.  The lattice spacing $b$ and box sizes $L$ are chosen so that $2\pi/L\ll p \ll \pi/b$  for each momentum $p$.

The strength of the triplet-channel coupling $c_t$ on the lattice is tuned to generate a bound deuteron with  energy $-2.225$ MeV.  This sparse matrix lattice eigenvector calculation also gives the bound state wave function $\psi_B(\bm{r})$. 
For the spin-singlet channel we use L\"{u}scher's finite volume method
\cite{Luscher:1986pf} to determine the interaction coefficient $c_s$.   L\"{u}scher's finite volume formula relates the two-particle energy
levels in a periodic cube to the scattering phase shift.   We use L\"{u}scher's method to tune the lattice 
coupling to produce the neutron-protron scattering length, $a=-23.71$ fm. 

In the continuum limit we can analytically calculate the dependence on the parameter $i\epsilon$ at infinite volume.  The exact result is a
 generalization of Eq.~(\ref{eq:EFT})
\begin{align}\label{eq:deltaEFT}
\mathcal M_C(\epsilon)=&\frac{1}{p^2+\gamma^2}-\frac{1}{(1/a+i p_\epsilon)(\gamma -i p_\epsilon)}\, ,
\end{align}
with $p_\epsilon=\sqrt{p^2+iM\epsilon}$.  The parameter $\epsilon$ serves as an infrared regulator which exponentially suppresses the contribution from scattering at large distances.  This is a key fact which we use to remove finite volume errors from the lattice calculations of $\mathcal M$ in Eq.~(\ref{eq:Greens}).  For nonzero $\epsilon$ the finite volume error becomes exponentially small as a function of box length $L$.  This allows us to perform lattice calculations for values of $L$ where the finite volume error is negligibly small.  We then use the fact that the scattering amplitude is smooth in $\epsilon$ to extrapolate to $\epsilon = 0$ using a simple linear extrapolation.  We note that there are computational methods such as the Lorentz Integral Transform method \cite{Efros:2007nq} which also compute the amplitude as a function of $\epsilon$.  However the simple extrapolation method appears to be the most efficient approach for the two-body capture problem we analyze here.   A similar complex energy method was used in Ref.~\cite{Kamada:2003xy} for Faddeev-Yakubovski calculations. 

\begin{figure}[thb]
\begin{center}
\includegraphics[width=0.47\textwidth,clip=true]{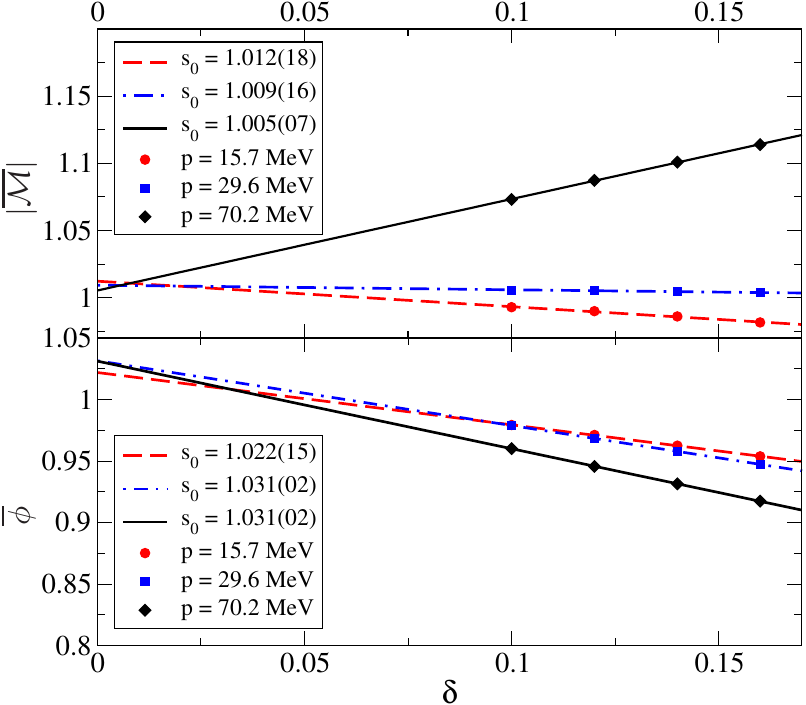} 
\end{center}
\caption{\protect  Linear extrapolation of lattice results (solid curves).  The top panel shows $|\overline{\mathcal M}|= |\mathcal M(\delta)|/ |\mathcal M_C(\delta=0)|$, and the bottom panel shows $\overline{\phi}= \arg[M(\delta)]/ \arg[M_C(\delta=0)].$
  }
\label{fig:dataB}
\end{figure}

For convenience we define the parameter $\delta = \epsilon M p^2$.  In our lattice calculations we compute the scattering amplitude at volumes where the finite volume error for our chosen value of $\epsilon$ is negligible.  We then extrapolate the lattice data to the $i\epsilon\rightarrow i0^+$ limit using the linear fit $s_0+s_1\,\delta$.  This extrapolation corresponds with taking the infinite volume limit.   The extrapolation for several values of $p$ are shown  in Fig.~\ref{fig:dataB}.  
In the top panel we show results for $|\overline{\mathcal M}|$ which is the ratio of $|\mathcal M(\delta)|$ to the continuum limit result at $\delta = 0$, $|\mathcal M_C(0)|$.  In the bottom panel we show results for $\overline{\phi}$ which is the ratio of $arg[\mathcal M(\delta)]$ to the continuum limit result at $\delta = 0$, $arg[\mathcal M_C(0)]$.

The errors of the extrapolation fit are indicated by the $s_0$ values in  the legends in Fig.~\ref{fig:dataB}. 
The deviation of the extrapolated value of $|\overline{\mathcal M}|$ and $\overline{\phi}$ from the value $1$ is a measure of the lattice spacing error.  The extrapolated lattice numbers agree with the continuum results for $|\mathcal M|$ to within $2\%$ and for $\phi$ to within 4\%.  We don't present results for other lattice spacings, but the comparison with the exact continuum limit results makes it clear that the lattice discretization errors are under good control.  

\begin{figure}[h]
\begin{center}
\includegraphics[width=0.47\textwidth,clip=true]{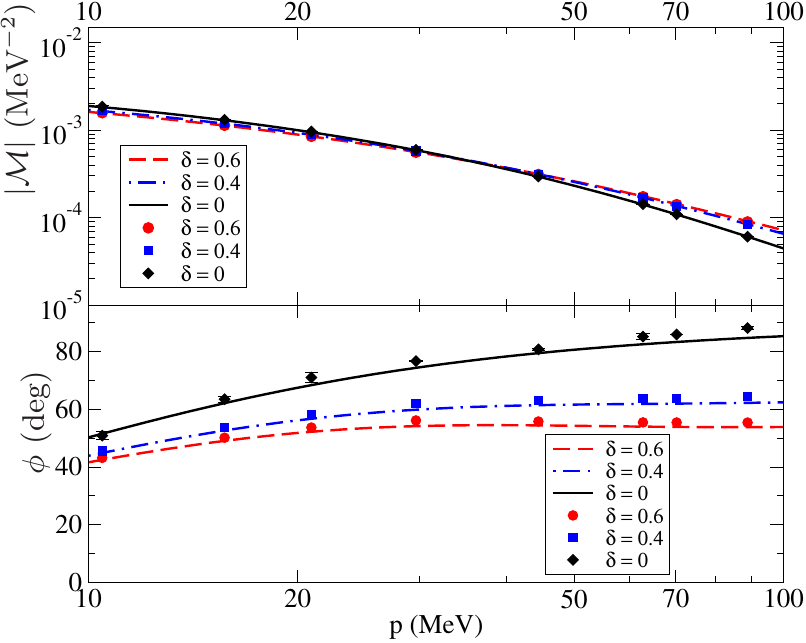} 
\end{center}
\caption{\protect  
Comparison of lattice results with continuum results from Eq.~(\ref{eq:deltaEFT}) for $\delta = 0.6, 0.4$ and results extrapolated to $\delta = 0$.  In the top panel we show $|\mathcal M|$ and in the bottom panel we show the phase angle $\phi$ in degrees.}
\label{fig:dataC}
\end{figure}

In Fig.~\ref{fig:dataC}  we show the lattice results from Eq.~(\ref{eq:Greens}) in comparison with the continuum results from Eq.~(\ref{eq:deltaEFT}) for $\delta = 0.6, 0.4$ and results extrapolated to $\delta = 0$.  In the top panel we show $|\mathcal M|$ and in the bottom panel we show the phase angle $\phi$ in degrees.  We see that the lattice results reproduce the continuum results in all cases with errors no more than a few percent.  This residual error can be attributed to lattice spacing discretization effects.

\emph{Discussion.}---
We have proposed a general formalism for \textit{ab initio} calculations of radiative capture cross sections on the lattice.  The first part of method requires performing  lattice simulations to determine the matrix elements of a multi-channel adiabatic Hamiltonian, reducing the incoming state to an effective two-body system and the outgoing nucleus to a single body.  The second part of the method uses the two-point Green's function for this adiabatic Hamiltonian to calculate the radiative capture cross section.

The numerical details of the first part of the method will be discussed in  forthcoming publications.  In this study we have demonstrated the second part of the method using the example of radiative neutron capture reaction $p(n,\gamma)d$ at leading order in pionless EFT.  We have shown that extrapolations in the parameter $\delta$ are an effective way to remove finite volume errors in the lattice calculations.  We were able to reproduce the known continuum result for the capture amplitude at infinite volume with an error of within $2\%$ for the magnitude and an error of within $4\%$ for the phase angle.  These remaining errors are consistent with the size of the small discretization errors due to lattice spacing.

This method should be applicable for calculating photonuclear reaction rates involving many nuclear systems.  This includes halo nuclei with a single valence nucleon.  For example, in $^{14}$C$(n,\gamma)^{15}$C the carbon-15 nucleus is a halo nucleus with a neutron separation energy of only $1.22$ MeV. At low energy the incoming state can be treated as a point-like neutron and a tightly-bound carbon-14 core in halo EFT~\cite{Rupak:2012cr}.
But \emph{ab initio} lattice calculations should be able to go well beyond halo systems.  This method could also be applied to more complicated systems such as alpha capture on carbon-12 or oxygen-16 as occurs in helium burning.  We are hopeful that such systems will studied in the near future using methods outlined here. 

\emph{Acknowledgments.}--- The authors thank E. Epelbaum, R. Higa, H. Krebs, T. L{\"a}hde, U.-G. Mei{\ss}ner for useful discussions.  G.R. thanks R. Higa for valuable discussions on the formulation of the Green's function approach. Computing support was provided by the HPCC at MSU. Part of this work was completed at the INT. Partial support provided by the U.S. Department of Energy grant DE-FG02- 03ER41260 for D.L. and  the U.S. NSF Grant No. PHY-0969378 for G.R.

\input{npdgamma_lattice_021813.bbl}

\end{document}

%% file: npdgamma_lattice_021813.bbl
%